\documentclass[aps,prl,reprint,superscriptaddress,floatfix,fleqn]{revtex4-2}

\usepackage[utf8]{inputenc}
\usepackage[T1]{fontenc}
\usepackage[main=english]{babel}

\usepackage[a4paper,margin=2cm]{geometry}

\usepackage{anyfontsize}
\usepackage[varvw]{newtx}
\usepackage[nopatch=footnote]{microtype}

\usepackage[dvipsnames]{xcolor}
\usepackage[hidelinks,colorlinks=true,allcolors=Blue]{hyperref}

\usepackage{mathtools, dutchcal, tensor, braket}
\let\t\tensor
\usepackage{braket}

\def\e{\mathrm e}	
\def\i{\mathrm i}	
\def\ii{\mathrm i}	
\def\d{\mathrm d}	
\def\dd{\mathrm d}	
\def\p{\partial}	

\def\Metric{g}

\def\metric{l}

\def\llapse{U}

\def\Killing{\mathcal K}

\def\ERC{\textsc{erc}}
\def\etal{\textit{et al}}
\def\GRAVITES{\textsc{gravites}}

\def\IDOF{\textsc{idof}}
\def\IDOFs{\textsc{idof}s}

\def\PPN{\textsc{ppn}}

\def\QGI{\textsc{qgi}}

\def\SC{\textsc{sc}}
\def\WKB{\textsc{wkb}}

\def\A{\mathscr A}
\def\herm{\mathcal h}
\def\H{\mathcal H}

\def\D{\mathcal D\!}
\def\bD{\bar{D}\!} 
\def\Del{D\!}
\def\del{\nabla\!}
\def\E{\mathcal E}
\def\Riesz{\mathcal R}
\renewcommand{\hbar}{\hslash}

\def\barphi{\bar\phi}

\usepackage[capitalize]{cleveref}
\usepackage{orcidlink}
\usepackage{aas_macros}

\hypersetup{
	pdftitle={Internal dynamics and guided motion in general relativistic quantum interferometry},
	pdfsubject={},
	pdfauthor={Thomas B. Mieling}
}

\renewcommand{\paragraph}[1]{\textit{#1}—\ignorespaces}

\begin{document}
\title{Internal dynamics and guided motion in general relativistic quantum interferometry}
\author{Thomas B. Mieling~\orcidlink{0000-0002-6905-0183}}
\affiliation{University of Vienna, Faculty of Physics and Research Network TURIS, Boltzmanngasse~5, 1090 Vienna, Austria}
\begin{abstract}
  The coupling between internal degrees of freedom of quantum systems
  and their overall motion in an external gravitational field plays a
  central role in multiple extensions of Einstein’s equivalence
  principle to quantum physics.  While previous models of such effects
  were predominantly restricted to linearized gravity and often
  required quantum particles to follow prescribed world-lines, this
  letter shows how such phenomena can be understood using generally
  covariant semi-classical approximations in the framework of quantum
  field theory in curved space-times.  This method provides a
  unification and generalization of previously established results,
  but also predicts new effects such as an influence of internal
  energies on field amplitudes, as well as correction terms to the
  internal Schrödinger equation that give rise to Berry phases.
\end{abstract} 
\maketitle

Internal degrees of freedom (\IDOFs) play an important role in
gravitational quantum interferometry: In experiments originally
proposed by Marletto and Vedral \cite{2017PhRvL.119x0402M} as well as
Bose \etal.\ \cite{2017PhRvL.119x0401B}, the gravitational interaction
of two particles is expected to induce entanglement between their
path-correlated spin-degrees of freedom, and even for interferometers
in \emph{external} gravitational fields (where gravity is treated as a
“given” background) \IDOFs\ of the interfering particles are
important---sometimes even central---elements of the experimental
design.  For example, Anastopoulos and Hu predicted that systems with
multiple internal energy levels acquire a state-dependent phase shift
when propagating through a gravitational field
\cite{2018CQGra..35c5011A, 2021arXiv210612514A} and Zych \etal.\
predicted that internal dynamics of quantum systems are influenced by
gravitational time dilation \cite{2011NatCo...2..505Z}.  In a
Mach–Zehnder interferometer, for example, this latter effect leads to
a loss in interferometric visibility that would not be expected in
Newtonian gravity \cite{2021EPJC...81..928B,2024PhRvA.109f2207B}. Such
effects were shown to arise for massive spin-$1/2$ particles
\cite{2023CQGra..40w5014A}, atoms \cite{2019PhRvA.100e2116S},
electromagnetically bound composite systems
\cite{2023LNP..1017..491G}, as well as extended “light clocks”
\cite{2023AVSQS...5a4401B} and are also expected to arise in qubit
dynamics \cite{2025PhRvA.111a2411B} and distributed atomic ensembles
\cite{2025arXiv250919501F}.  This mechanism was suggested to be a
universal source of decoherence
\cite{2015NatPh..11..668P,2017NJPh...19b5011P}.

\IDOFs\ are also used in experiments demonstrating gravitationally
induced phase shifts, where particles are sent along specific
trajectories through a gravitational field \cite{2012PhRvL.108w0404H,
  2020PhRvX..10b1014R}. In atomic-fountain experiments, this is
achieved using optical pulses that drive Raman or Bragg transition
between internal states of the atoms \cite{1991PhRvL..67..181K,
  1992ApPhB..54..321K, 2013NJPh...15b3009A}. Detailed modeling of
these transitions is necessary to correctly interpret the overall
phase shift observed in such interferometers
\cite{2011CQGra..28n5017W, 2021AmJPh..89..324O, 2012Giulini,
  2025PhRvA.112f2211M}.  The interplay between \IDOFs\ and guidance in
external gravitational fields is even more pronounced in the recent
“Quantum Galileo interferometer” (\QGI) experiment, which used
different atomic states in its two interferometer arms: one with a
magnetic moment that allowed the atoms to be levitated in an
inhomogeneous magnetic field, and one without such a magnetic moment
so that the atoms were falling freely \cite{2025arXiv250214535D}.  In
this experiment, the overall phase shift is fully determined by the
semi-classical trajectories forming the two interferometer arms
\cite{2019NJPh...21d3047H,2020FrP.....8..176M}, but the equations
governing these world-lines depend on the \IDOFs\
\cite{2025arXiv250415409A,2025arXiv250421626D}.

The dynamics of composite systems also play a significant role in
extensions of various equivalence principles from classical physics
\cite{1977AmJPh..45..903O} to quantum physics, such as the
proportionality of mass and rest-energy
\cite{2018NatPh..14.1027Z,2019PhRvA.100e2116S,2022PhRvA.106e2801T};
the identity of inertial and gravitational masses
\cite{2017NatCo...815529R,2025arXiv250214535D}; the universality of
free fall of test particles \cite{2020FrP.....8..176M}; or
combinations thereof \cite{2016CQGra..33sLT01O}, see also
Refs.~\cite{2012Giulini,2019SciA....5.8966L,2023LNP..1017..491G}.
However, current analyses of the aforementioned effects involving
\IDOFs\ in gravitational quantum physics suffer from multiple
deficits.  Firstly, to demonstrate novel gravitational effects, the
setups described in Refs.~\cite{2012PhRvL.108w0404H,
  2017NJPh...19b5011P} require guiding potentials, but their influence
on the particle’s phase was left unmodeled. Relativistic descriptions
of such effects were developed in Ref.~\cite{2020PhRvX..10b1014R}, but
to describe state-dependent guidance as used in the \QGI\ experiment,
this requires making the \textit{ad-hoc} assumption that the guiding
potential are state-dependent.  Additionally, the majority of current
models are limited to weak gravitational fields, often assuming the
validity of the linearized Einstein equations. Without further
analyses, the resulting predictions thus cannot be used in the
parametrized post-Newtonian framework (\PPN\
\cite{2014LRR....17....4W}) to assess which specific aspects of the
gravitational field are relevant for the experiments and how well
(future) experimental results agree with general relativity, as
compared to alternative theories of gravity.  Finally, there is, so
far, no general model that puts the various effects described above
into a coherent picture.  For example the model of Anastopoulos and Hu
\cite{2018CQGra..35c5011A,2021arXiv210612514A} accounts for
gravitationally induced phase shifts that depend on the internal
energy but assumes internal dynamics to be trivial, while the model of
Zych \etal.\ \cite{2011NatCo...2..505Z} models \IDOF-dynamics but does
not describe phase-effects as predicted by Anastopoulos and Hu.

The purpose of this letter is to point out that these gaps can be
filled by applying semi-classical approximation methods to a
phenomenological field theory based on the equation
\begin{align}
	\label{eq:model}
	(\hbar c)^2 \t\Metric{^\mu^\nu} \t\D{_\mu} \t\D{_\nu} \phi
		= \E^2 \phi,
\end{align}
where $\phi = (\t\phi{^a})$ is a \emph{multi-component} field whose
number of components, $N$, is chosen as the number of \IDOFs, and
$\E = (\t\E{^a_b})$ is a positive Hermitian $N \times N$ matrix that
is chosen to model the energies and internal dynamics of the internal
states (details are described below).  In this equation $\hbar$ is the
reduced Planck constant, $c$ is the speed of light in vacuo,
$\t\Metric{^\mu^\nu}$ is the contravariant space-time metric
(describing the gravitational field),
$\t\D{_\mu} = \t\Del{_\mu} - \ii (q / \hbar) \t A{_\mu}$ is a
gauge-covariant derivative, in which $\t\Del{_\mu}$ is the Levi-Civita
derivative of $\t\Metric{_\mu_\nu}$, $q$ is the electric charge, and
$\t A{_\mu}$ is an external electromagnetic potential.

\paragraph{Non-relativistic limit}
The limit $c \to \infty$ of \cref{eq:model} was already studied in
Appendix~A of Ref.~\cite{Zych2017}, Sec.~III.A of
Ref.~\cite{2021EPJC...81..928B}, and Sec.~2.6 of
Ref.~\cite{2021arXiv210612514A}. While these works assumed $\E$ to be
constant, \cref{eq:model} also allows for $\E$ to vary with space and
time. As is explained below, this makes it possible to describe guided
motion in external potentials.  In either case, the non-relativistic
limit is based on a decomposition of $\E$ in form $m c^2 I + \hat\H$,
where the first term, proportional to the identity matrix, describes
the rest-mass energy and $\hat\H$ is a non-relativistic internal
Hamiltonian. The Newtonian limit then yields the Schrödinger equation
\begin{align}
	\label{eq:Schrödinger}
		\ii \hbar \frac{\p}{\p t} \Psi
			={}& \frac{1}{2m} \t\metric{^i^j} \t{\hat p}{_i} \t{\hat p}{_j} \Psi
			+ m \llapse \Psi
			+ q V \Psi
			+ \hat\H \Psi
			,
\end{align}
where $\t l{^i^j}$ is the contravariant spatial metric with
Levi-Civita derivative $\t\del{_i}$,
$\t{\hat p}{_i} = - \ii \hbar \t\nabla{_i} - q \t A{_i}$ is the
spatial momentum operator, $\llapse$ is the Newtonian gravitational
potential, and $V = - \t A{_t}$ is the scalar electromagnetic
potential (a generalized equation applies in references frames that
are stationary but not static, see the supplementary materials
\cite{supplement}).  Due to the complexity of solving the
three-dimensional Schrödinger in realistic potentials, quantum
interferometers in Newtonian gravitational fields are commonly
analyzed using semi-classical approximations based on \WKB\ methods
\cite{1996GReGr..28.1043L, 1974PhRvL..33.1237O, 2022Sci...375..226O,
  2025arXiv250214535D}. However, the \WKB\ approximation can also be
applied to the general-relativistic equation \eqref{eq:model} without
taking a Newtonian limit. As is demonstrated below, the resulting
semi-classical equations are not significantly more complicated than
their non-relativistic counterparts (this extends the analysis of
unguided motion without \IDOFs\ provided in
Ref.~\cite{2025PhRvA.112f2211M}).

\paragraph{Semi-classical limit}
In the limit $\hbar \to 0$, the field $\phi$ is described by a phase
function $S$ (with dimension of action), a scalar amplitude $\A$, and
an internal state vector $\psi$, all of which are functions of space
and time.  As is derived explicitly in the supplementary material
\cite{supplement}, if the energy matrix has the form
$\E = \E_\SC + \hbar \varOmega + O(\hbar^2)$, the \WKB\ approximation
to \cref{eq:model} implies the following: The internal state vector
$\psi$ satisfies the eigenvalue equation
\begin{align}
	\E_\SC \psi = E \psi,
\end{align}
where both $\E_\SC$ and $E$ may depend on space and time.  The
corresponding mass $M = E/c^2$ determines the eikonal equation
\begin{align}
	\t\Metric{^\mu^\nu} \t p{_\mu} \t p{_\nu} = - (M c)^2,
\end{align}
where $\t p{_\mu} = \t\Del{_\mu} S - q \t A{_\mu}$ is the
four-momentum.  Moreover, the phase function $S$, amplitude $\A$, and
internal state vector $\psi$ satisfy the following transport equations
along the four-velocity
$\t u{^\mu} = \t\Metric{^\mu^\nu} \t p{_\nu} / M$:
\begin{subequations}
\begin{align}
	\label{eq:transport:S}
	\t u{^\mu} \t\Del{_\mu} S
	&= - M c^2 + q \t A{_\mu} \t u{^\mu},
	\\
	\label{eq:transport:A}
	\t u{^\mu} \t\Del{_\mu} \A
	&= - \frac{\t\Metric{^\mu^\nu} \t\Del{_\mu} \t p{_\nu}}{2M} \A,
	\\
	\label{eq:transport:psi}
	\ii \t u{^\mu} \t\bD{_\mu} \psi
		&= \varOmega \psi + [\Riesz, \varOmega] \psi.
\end{align}
\end{subequations}
Here, $\Riesz$ is the Riesz projector onto the eigenspace of $\E_\SC$
associated to the eigenvalue $E$ and $\t\bD{_\mu}$ is the associated
Berry connection, acting as
$\t\bD{_\mu} \psi = \Riesz \t\Del{_\mu} \psi = \t\Del{_\mu} \psi -
(\t\Del{_\mu} \Riesz) \psi$.

Decomposing $E$ into a constant bare-mass energy and an interaction
energy as $E = m c^2 + w$, the four-force
$\t f{_\mu} = \t u{^\nu} \t\Del{_\nu} \t p{_\mu}$ takes the form
\begin{align}
	\label{eq:force}
	\t f{_\mu} = - \t\Del{_\mu} w + q \t F{_\mu_\nu} \t u{^\nu},
\end{align}
with
$\t F{_\mu_\nu} = \t\Del{_\mu} \t A{_\nu} - \t\Del{_\nu} \t A{_\mu}$
denoting the electromagnetic field strength, so $w$ can be thought of
as a state-dependent interaction potential.  Hence, every integral
curve $\gamma$ of $\t u{^\mu}$ satisfies classical equations of
motion. \Cref{eq:transport:S,eq:transport:A,eq:transport:psi} then
imply evolution equations along such world-lines.  Specifically,
\cref{eq:transport:S} implies that the quantum phase $S / \hbar$
evolves as
\begin{align}
	\label{eq:dS / dt}
	\dd S / \hbar
		= - \omega_\textsc{c}\, \dd \tau
		+ (w / \hbar) \dd \tau
		+ (q / \hbar) \t A{_\mu} \t{\dd x}{^\mu},
\end{align}
where $\tau$ is the proper time along $\gamma$ and
$\omega_\textsc{c} = m c^2 / \hbar$ is the Compton frequency.  The
first term describes the well-known fact that the phase of free
particles evolves with proper-time at the rate of the Compton
frequency \cite{2010Natur.463..926M, 2011CQGra..28n5017W,
  2012CQGra..29d8001H, 2012CQGra..29d8002W, 2011CQGra..28n5018S,
  2012Giulini, 2025PhRvA.112f2211M}.  If $w$ depends on the internal
state but is constant in space and time, the second term implies phase
shifts as first predicted in Refs.~\cite{2018CQGra..35c5011A,
  2021arXiv210612514A}. Alternatively, a non-constant $w$ can arise
from external potentials: phase shifts of this kind are particularly
well known in the Newtonian regime \cite{1979RvMP...51...43G}. In the
relativistic context, on the other hand, the explicit form of such
phase contributions along individual world-lines was first suggested
in Ref.~\cite{2020PhRvX..10b1014R}.  Finally, the last term in
\cref{eq:dS / dt} accounts for phases induced by the electromagnetic
field.  The dependence of this last term on the potential $\t A{_\mu}$
instead of field-strength $\t F{_\mu_\nu}$, which determines the
applied force \eqref{eq:force}, leads to the Aharonov–Bohm effect
\cite{1949PPSB...62....8E,1959PhRv..115..485A}.  The second term,
$(w / \hbar) \d\tau$, gives rise to similar effects that can occur
even for uncharged particles \cite{1984PhRvL..53..319A,Zeilinger1986}.
It is thus natural to ask whether analogous effects can be induced by
the term $- \omega_\textsc{c} \dd\tau$.  However, due to the lack of
an associated force term in \cref{eq:force}---which is related to the
fact that Einsteinian gravity has no invariant notion of a
gravitational force--- and the constancy of $\omega_\textsc{C}$, a
universally agreed-upon definition of a gravitational Aharonov–Bohm
effect in general relativity is still lacking.
Refs.~\cite{Zeilinger1986,2012PhRvL.108w0404H} introduced such a
notion in the context of linearized gravity (where a flat background
metric and a preferred reference frame can be used to give meanings to
the terms “gravitational potential” and “gravitational force”) and the
observation of a related phase shift was reported in
Ref.~\cite{2022Sci...375..226O}. However, the notion of a
gravitational Aharonov–Bohm effect used there differs significantly
from that used in the theoretical literature
\cite{1981JPhA...14.2353F,1983JPhA...16.2457A,1990AnPhy.203..392B}.

\Cref{eq:transport:A}, which is equivalent to
$\t\Del{_\mu}(\A^2 M \t u{^\mu}) = 0$, implies the following evolution
of the amplitude $\A$ with respect to proper time:
\begin{align}
	\label{eq:dA / dt}
	\frac{1}{\A}
	\frac{\dd \A}{\dd \tau}
		&= -
		\frac{1}{2} \left(
		\t\Del{_\mu} \t u{^\mu}
		+ \frac{1}{m c^2 + w} \frac{\dd w}{\dd \tau}
		\right).
\end{align}
The loss in amplitude associated with the ray divergence
$\t\Del{_\mu} \t u{^\mu}$ is well known in ray optics
\cite{KravtsovOrlov1990,2021CQGra..38q5007M} and also arises in
semi-classical dynamics of massive fields
\cite{1981JPhA...14..411A,2023PhRvD.107d4029O}.  The additional
contribution arising from changes in the energy $w$ appears to be a
new result.

Finally, decomposing the internal state vector $\psi$ in an adapted
frame (orthonormal eigenvectors of $\E_\SC$ with eigenvalue $E$),
\cref{eq:transport:psi} can be written in the form of a Schrödinger
equation,
\begin{align}
	\label{eq:dpsi / dt}
	\ii \hbar \frac{\dd \t\psi{^\alpha}}{\dd \tau}
	= \t\H{^\alpha_\beta} \t\psi{^\beta}
	- \hbar \t\varpi{^\alpha_\beta_\mu} \t\psi{^\beta} \t u{^\mu},
\end{align}
where $\t\H{^\alpha_\beta}$ are the matrix elements of the internal
Hamiltonian $\H = \hbar \varOmega$ in the adapted frame, and
$\t\varpi{^\alpha_\beta_\mu}$ are the Berry connection coefficients,
see the supplementary materials for details \cite{supplement}.  That
\IDOFs\ should obey a Schrödinger equation with proper time as the
evolution parameter was first suggested in
Ref.~\cite{2011NatCo...2..505Z}. The present analysis shows that this
is a direct consequence of the field equation \eqref{eq:model}.  The
influence of a Berry connection on internal dynamics appears to have
been overlooked in the field of gravitational quantum interferometry,
despite the fact that such terms are known to enter semi-classical spin
dynamics derived from Dirac’s equation
\cite{2007PhLA..368..356G,2023PhRvD.107d4029O}.

\paragraph{Field quantization}
Whereas previous models describing the phase and internal dynamics
along individual world-lines had to postulate explicit formulas
relating quantum-theoretical transition amplitudes to phase shifts and
inner products of internal state vectors \cite{2011NatCo...2..505Z},
such additional assumptions are not necessary in the present
field-theoretic setting. Indeed, the field equation \eqref{eq:model}
can be treated using standard methods of quantum field theory in
curved space-times, according to which all transition probabilities in
a stationary space-time with a distinguished vacuum state $\ket 0$ are
expressible in terms of Klein–Gordon products
\cite{1975RSPSA.346..375A,2015PhR...574....1H}.  For
positive-frequency solutions $\phi'$ and $\phi''$ one has
\begin{multline}
	\hspace{-1em}
	\braket{0 | a(\barphi') a^\dagger(\phi'')  | 0}
		= \braket{0 | [\Phi(\barphi'), \Phi^\dagger(\phi'')]  | 0}
		\\
		= \frac{\i}{\hbar c} \!\! \int_\Sigma \hspace{-0.5em} \sqrt{-\Metric} \t\Metric{^\mu^\nu} [
			\herm(\t\D{_\mu} \barphi', \phi'')
			- \herm(\barphi, \t\D{_\mu} \phi'')
		]  \t{\d\Sigma}{_\nu},
\end{multline}
where $a$ and $a^\dagger$ are ladder operators, $\Phi$ and
$\Phi^\dagger$ are quantum field operators, $\Sigma$ is any Cauchy
surface, and $\herm(\barphi', \phi'')$ denotes the point-wise inner
product of $\phi'$ and $\phi''$. If the fields have the asymptotic
form $\phi \sim (\hbar / \sqrt{2m}) \A \psi \, \e^{\i S / \hbar}$ as
$\hbar \to 0$, where the overall factor ensures that $\A^2$ has
dimensions of a density, the two-point function
$\braket{0 | a(\barphi') a^\dagger(\phi'') | 0}$ has the asymptotic
behavior
\begin{multline}
	\braket{0 | a(\barphi') a^\dagger(\phi'')  | 0}
		\\
		\sim
		\frac{1}{2m c}
    \int_\Sigma
			\sqrt{-\Metric}
			\bar\A' \A''
			\herm(\bar\psi', \psi'')
			\\\times
			\t\Metric{^\mu^\nu}
			(\t*p{^\prime_\mu} + \t*p{^\prime^\prime_\mu} )
			\exp[-\i(S' - S'') / \hbar]
		\t{\d\Sigma}{_\nu}.
\end{multline}
According to the principle of stationary phases, this transition
amplitude vanishes if the phase difference $\varDelta S = S' - S''$
has no stationary points within the joint support of $\A'$, $\A''$,
$\psi'$, and $\psi''$.  The main alternative of interest is
$\varDelta S$ being constant, in which case the relevant transition
amplitude takes the form
\begin{multline}
	\label{eq:transition amplitude}
	\hspace{-1em}
	\braket{\A', \psi', S' \mid \A'', \psi'', S''}
		\\
		= \e^{- \ii \varDelta S / \hbar} 
		\int_\Sigma \hspace{-0.3em} \bar\A' \A'' \herm(\bar \psi', \psi'') \varrho \, \dd V,
\end{multline}
where $\dd V$ is the volume element associated to stationary observers
whose four-velocity $\t{\mathring u}{^\mu}$ is proportional to the
timelike Killing vector field $\t\Killing{^\mu}$, and $\varrho$ is a
scalar weight depending on the energy flux through the considered
time-slice. Explicitly, one has
\begin{align}
	\varrho
		= \left( 1 + \frac{w}{m c^2} \right)
		\frac{\t u{^\mu} \t n{_\mu}}{\t{\mathring u}{^\nu} \t n{_\nu}},
\end{align}
where $\t n{_\mu}$ is co-normal to the Cauchy surface of integration,
see the supplementary material for a derivation \cite{supplement}.  In
practical applications, external potentials are weak compared to the
rest-mass energy, and particle velocities in interferometers are small
when compared to $c$, which leads to the approximation
$\varrho \approx 1$.  Furthermore, if $\herm(\bar \psi', \psi'')$
varies slowly across the support of the interfering wave packets, this
almost constant factor can be pulled out of the integral in
\cref{eq:transition amplitude}.  Finally, if the amplitudes of the
wave packets can be approximated by Gaussian distributions that are
sufficiently narrow that the curvature of the spatial geometry is
negligible, one obtains the approximation formula
\begin{multline}
	\label{eq:transition amplitude approximate}
	\hspace{-1em}
	\braket{\A', \psi', S' \mid \A'', \psi'', S''}
	\\
		\hspace{-1em}
		\approx \exp(- \ii \varDelta S / \hbar)
		\,
		\herm(\bar \psi', \psi'')
		\,
		\sqrt[4]{\det(\Upsilon' \Pi^2 \Upsilon'')}
		\\\times
		\exp[
			- \tfrac{1}{8} \t{\varDelta x}{^i} \t*\Upsilon{^\prime_i_j} \t\Pi{^j^k} \t*\Upsilon{^\prime^\prime_k_l} \t{\varDelta x}{^l}
		],
\end{multline}
where $\t{\varDelta x}{^i}$ is the spatial separation between the
centers of the two Gaussian distributions (in spatial Riemannian
coordinates), $\t\Upsilon{^\prime}$ and $\t\Upsilon{^\prime^\prime}$
are the inverse covariance matrices of the distributions, and $\Pi$ is
the matrix inverse to
$\frac{1}{2} (\t\Upsilon{^\prime} + \t\Upsilon{^\prime^\prime})$.  For
$\t{\varDelta x}{^i} = 0$ and
$\t\Upsilon{^\prime} = \t\Upsilon{^\prime^\prime}$ this reproduces the
results of Ref.~\cite{2011NatCo...2..505Z}.  The separation-induced
loss in interferometric visibility (which arises whenever
$\t{\varDelta x}{^i} \neq 0$) is analogous to that predicted for
photons in Ref.~\cite{2012CQGra..29v4010Z}.  \Cref{eq:transition
  amplitude approximate} also implies that losses in visibility can
result from differences in deformation of interfering wave packets
that arise from the propagation through an interferometer. The
geodesic deviation equation (see, e.g., Sec.\ 3.3 and 9.2 of
Ref.~\cite{Wald_1984}) implies that such deformations of wave packets
can occur even in free fall.

\paragraph{Applications}
The simplest case of \cref{eq:model} is obtained by setting $N = 1$.
While this case neglect internals dynamics, it nonetheless models
guided motion through a potential $w$. \Cref{eq:force,eq:dS / dt} then
describe the force on the world-lines and the associated phase
shifts. These relativistic equations were first formulated in a
non-field-theoretic setting in Ref.~\cite{2020PhRvX..10b1014R}.

Guiding potentials may, however, depend on the internal state of the
system. For example, an uncharged particle with internal states having
different dipole tensors $\t\mu{^\rho^\sigma}$ can be modeled by
setting
\begin{align}
	\t\E{_\SC^a_b}
		= m c^2 \t*\delta{^a_b}
		+ \t\mu{^a_b^\rho^\sigma} \t F{_\rho_\sigma},
\end{align}
or in an eigenbasis,
\begin{align}
	\E_\SC
		=
		\begin{pmatrix}
			mc^2 + \t*\mu{_1^\rho^\sigma} \t F{_\rho_\sigma} & 0
			\\
			0 & mc^2 + \t*\mu{_2^\rho^\sigma} \t F{_\rho_\sigma}
		\end{pmatrix}.
\end{align}
Phenomenologically, a spin-$1/2$ particle can thus be modeled by
setting the diagonal elements to
$\pm \t*\mu{^\nu^\rho} \t F{_\nu_\rho}$, corresponding to parallel or
antiparallel alignment relative to an external field.  In the \QGI\
experiment, the two relevant states are $\ket{F = 2, m_F = 1}$ and
$\ket{F = 1, m_F = 0}$ of \textsuperscript{87}Rb
\cite{2025arXiv250214535D}, which correspond to
$\t*\mu{_1^\rho^\sigma} \t F{_\rho_\sigma} = + \mu B$ and
$\t*\mu{_1^\rho^\sigma} \t F{_\rho_\sigma} = 0$, where $\mu$ is the
atom’s magnetic moment and $B$ is the external magnetic field
strength.

Internal dynamics of freely falling particles are obtained by setting
$\E_\SC = m c^2 I$ and allowing for general internal Hamiltonians
$\H$. While previous analyses have mostly assumed $\H$ to be constant,
the present formalism allows for internal Hamiltonian that depend on
space and time.

\paragraph{Conclusion}
When combined with semi-classical approximations, the quantum field
theory based on \cref{eq:model} provides a fully relativistic but
nonetheless analytically tractable model for current and currently
proposed interference experiments involving particles with internal
degrees of freedom coupled to guiding potentials, as well as external
electromagnetic and gravitational fields.

Previous investigations of the phenomena described by the evolution
equations \eqref{eq:dS / dt}, \eqref{eq:dA / dt}, and \eqref{eq:dpsi /
  dt} derived these effects in isolation. The present analysis shows
how these effect fit together: they arise consistently from the
semi-classical approximation to the field equation
\eqref{eq:model}. This derivation also provides a significant
extension of the range of applicability of these equations: they are
not limited to weak gravitational fields, nor are they restricted to
individual world-lines. Instead, these equations describe
\emph{fields} in arbitrary space-times. As such, their quantization
can be carried out according to established methods without the need
for further postulates. When applied to stationary space-times, where
the notion of individual particles has a clear meaning, the resulting
framework is not limited to single-particle interference: using Wick’s
theorem, multi-particle transition amplitudes can be expressed in
terms of products of single-particle amplitudes that can be computed
using \cref{eq:transition amplitude}.  By allowing for complex fields,
this model is not limited to ordinary matter but also describes
antimatter, whose gravitational coupling is studied by the ALPHA-g
experiment \cite{2023Natur.621..716A}.  Such comparisons between
free-fall behaviors of matter, antimatter, and composite systems can
provide further tests of the weak equivalence principle
\cite{1987_Hughes_Goldman_Nieto,2025NJPh...27i5001M}.

The present model can be used to describe particles with spin on a
phenomenological basis. More rigorous descriptions can be obtained by
allowing for spinorial fields instead, in which case the Berry
connection naturally depends on the space-time geometry
\cite{2023PhRvD.107d4029O,2026_01_Oancea}.  In such cases, internal
dynamics of particles are influenced not only by time dilation, but
also by the spin structure associated to the gravitational field.  But
even without modifications, the present model allows for further
investigations into the behavior of particles with \IDOFs\ in external
fields. For example, the equations derived here do not require
Einstein’s field equations to hold, which makes it possible to study
the influence of potential violations of these equations using the
\PPN\ formalism.

\paragraph{Acknowledgments}
This research is funded by the European Union (\ERC, \GRAVITES,
project no.~101071779). Views and opinions expressed are however those
of the authors only and do not necessarily reflect those of the
European Union or the European Research Council Executive
Agency. Neither the European Union nor the granting authority can be
held responsible for them.

\bibliography{bibliography}
\end{document}


\title{Supplementary Material:\\
Internal dynamics and guided motion in general relativistic quantum interferometry}
\maketitle

\section{Stationary space-times}

In stationary space-times, locally there exist coordinates
$(\t x{^\mu}) = (t, \t x{^i})$ relative to which the metric tensor
takes the form
\begin{align}
	\label{eq:metric decomposition}
	\Metric
		= - \lapse^2 (c \dd t - \t\shift{_i} \t{\dd x}{^i})^2
		+ \t\metric{_i_j} \t{\dd x}{^i} \t{\dd x}{^j},
\end{align}
in which the lapse $\lapse$, shift $\t\shift{_i}$, and spatial metric
$\t\metric{_i_j}$ are independent of the temporal coordinate $t$.  A
detailed discussion of the geometric interpretation of these
quantities is provided in Sec.~II of Ref.~\cite{2025PhRvA.112f2211M}.
The contravariant metric components, $\t\Metric{^\mu^\nu}$, and the
induced density, $\sqrt{-\Metric}$, take the form
\begin{subequations}
\begin{align}
	c^2 \t\Metric{^0^0}
		&= \t\metric{^i^j} \t\shift{_i} \t\shift{_j} - \lapse^{-2},
	&
	\t\Metric{^i^j}
		&= \t\metric{^i^j},
	\\
	c \t\Metric{^0^i}
		&= \t\metric{^i^j} \t\shift{_j},
	&
	\sqrt{-\Metric}
		&= c \lapse  \sqrt{\metric},
\end{align}
\end{subequations}
where $\t\metric{^i^j}$ is the inverse of $\t\metric{_i_j}$ and
$\sqrt\metric$ is the density induced by the spatial metric.  The
parametrization in \cref{eq:metric decomposition} differs from the
\ADM\ decomposition
\begin{align}
	\label{eq:metric decomposition ADM}
	\begin{split}
		\Metric
			=& - c^2 \t*\lapse{_\ADM^2} \dd t^2
			\\&
			+ \t*\metric{^\ADM_i_j} (\t{\dd x}{^i} + c \t*\shift{_\ADM^i} \dd t) (\t{\dd x}{^j} + c \t*\shift{_\ADM^j} \dd t).
	\end{split}
\end{align}
Specifically, in terms of the \ADM\ quantities $\t*\lapse{^\ADM}$,
$\t*\shift{^\ADM_i}$ and $\t*\metric{^\ADM_i_j}$, the parameters
entering \cref{eq:metric decomposition} can be written as
\begin{subequations}
\begin{align}
	\t*\lapse{^2}
		&= \t*\lapse{_\ADM^2} - \t*\metric{_\ADM^i^j} \t*\shift{^\ADM_i} \t*\shift{^\ADM_j},
	\\
	\t*\shift{_i}
		&= \t*\shift{^\ADM_i} / \t*\lapse{^2},
	\\
	\t*\metric{_i_j}
		&= \t*\metric{^\ADM_i_j} + \t*\lapse{^2} \t*\shift{_i} \t*\shift{_j}.
\end{align}
\end{subequations}
The difference between these two parametrizations is due to
\cref{eq:metric decomposition} being adapted to a timelike Killing
vector field, while \cref{eq:metric decomposition ADM} is adapted to a
time function.  This paper uses \cref{eq:metric decomposition} instead
of \cref{eq:metric decomposition ADM}, as this choice leads to more
concise equations.

\section{Action and field equations}

The weighted bilinear action of the model, referred to as a
generalized Lagrangian in Sec.\ 4.1 of Ref.~\cite{Rejzner_2016}, is
given by
\begin{align}
	S_f[\varphi, \phi]
		= \int_\Mfd f L[\varphi, \phi]  \d\text{\textmu},
\end{align}
where $\Mfd$ is the space-time manifold, $f$ is a test function,
$\d\text{\textmu}$ is the metric volume element, which locally takes
the form
\begin{align}
	\d\text{\textmu}
		= (\sqrt{- \Metric} / c) \d^4 x,
\end{align}
and $L[\varphi, \phi]$ is the Lagrangian scalar 
\begin{align}
	\begin{split}
		L[\varphi, \phi] = &
			- \t\Metric{^\mu^\nu} \t\herm{_\a_b} \t\D{_\mu} \t\varphi{^\a}  \t\D{_\nu} \t\phi{^b}
			\\&\quad
			- (\hbar c)^{-2} \t\herm{_\a_b} \t\varphi{^\a} \t\E{^b_c} \t\E{^c_d} \t\phi{^d}.
	\end{split}
\end{align}
Here, $\phi$ and $\varphi$ are considered as sections of an
$N$-dimensional complex vector bundle and its complex conjugate,
respectively [both of physical dimension
$\sqrt{\dimf M \dimf L} / \dimf T$, so that $\varphi \phi$ has the
dimension of force].  Accordingly, indices $a$, $b$, $\ldots$ refer to
a local frame of a vector bundle while $\a$, $\b$, $\ldots$ refer to
the corresponding conjugate frame.  Assuming trivial bundles for
simplicity, the gauge-covariant derivatives arising here are taken to
be
\begin{subequations}
\begin{align}
	\t\D{_\mu} \t\varphi{^\a} &
		= \t\p{_\mu} \t\varphi{^\a} + \ii (q/\hbar) \t A{_\mu} \t\varphi{^\a},
	\\
	\t\D{_\mu} \t\phi{^a} &
		= \t\p{_\mu} \t\phi{^a} - \ii (q/\hbar) \t A{_\mu} \t\phi{^a},
\end{align}
\end{subequations}
where $\t A{_\mu}$ is the electromagnetic potential and $q$ is the
electric charge of the particles considered.  Finally,
$\t\herm{_\a_b}$ is a non-degenerate Hermitian matrix with constant
coefficients, and $\t\E{^a_b}$ is a matrix independent of $\phi$ and
$\varphi$ (whose components need not be constant).  The associated
Euler–Lagrange equations are
\begin{subequations}
\begin{align}
	\label{s:eq:field equation phi}
	(\hbar c)^2 \t\Metric{^\mu^\nu} \t\D{_\mu} \t\D{_\nu} \t\phi{^a}
		&= \t\E{^a_b} \t\E{^b_c} \t\phi{^c},
	\\
	\label{s:eq:field equation barphi}
	(\hbar c)^2 \t\Metric{^\mu^\nu} \t\D{_\mu} \t\D{_\nu} \t\varphi{^\a}
		&= \t\varphi{^\c} \t\E{_\c^\b} \t\E{_\b^\a},
\end{align}
\end{subequations}
where $\t\E{_\a^\b} = \t\herm{_\a_a} \t\E{^a_b} \t\herm{^b^\b}$, in
which $\t\herm{^a^\b}$ is the inverse of $\t\herm{_\a_b}$.  In the
following, it will be assumed that $\E$ is self-adjoint in the sense
that $\t\E{_\a^\b}$ is the complex conjugate of $\t\E{^b_a}$.  This
implies that solutions to \cref{s:eq:field equation barphi} can be
obtained from solutions to \cref{s:eq:field equation phi} by complex
conjugation.  As a consequence, it suffices to analyze
\cref{s:eq:field equation phi}, whose explicit form is given by
\begin{multline}
	(\t\p{_\mu} - \ii q \t A{_\mu}/\hbar)\left[
		\sqrt{- \Metric} \t\Metric{^\mu^\nu} (\t\p{_\nu} - \ii q \t A{_\nu} / \hbar) \t\phi{^a}
	\right]
	\\=
	\frac{\sqrt{- \Metric}}{(\hbar c)^2}\t\E{^a_b} \t\E{^b_c} \t\phi{^c}.
\end{multline}
The corresponding quantum theory can be formulated without using
explicit solutions to this equation.  Perturbative schemes yielding
appropriate solutions are described below.

\section{Klein–Gordon product}

On shell, the variation of the Lagrangian density
$\mathscr L = \sqrt{- \Metric} L / c$ is given by
\begin{align}
	\delta \mathscr L[\varphi, \phi]
		= \t\p{_\mu}\left(
			\delta\t\varphi{^\a} \frac{\p \mathscr L}{\p \t\p{_\mu}\t\varphi{^\a}}
			+ \frac{\p \mathscr L}{\p \t\p{_\mu}\t\phi{^a}} \delta\t\phi{^a}
		\right).
\end{align}
Because of the global $\mathrm{U}(1)$-symmetry of the model, one has
$\delta \mathscr L = 0$ for
$\delta\t\varphi{^\a} = + \i \t\varphi{^\a} / \hbar$ and
$\delta\t\phi{^a} = - \i \t\phi{^a} / \hbar$, which implies the
conservation law $\t\p{_\mu} \t j{^\mu} = 0$ for the vector density
\begin{multline}
	\t j{^\mu}[\varphi, \phi]
		=
			\frac{\i}{\hbar} \left(
				  \t\varphi{^\a} \frac{\p \mathscr L}{\p \t\p{_\mu}\t\varphi{^\a}}
				- \frac{\p \mathscr L}{\p \t\p{_\mu}\t\phi{^a}} \t\phi{^a}
			\right)
		\\
		= \frac{\i}{\hbar} \frac{\sqrt{-\Metric}}{c} \t\Metric{^\mu^\nu} [
			  \herm(\t\D{_\nu} \varphi, \phi)
			- \herm(\varphi, \t\D{_\nu} \phi)
		],
\end{multline}
where $\herm(\varphi, \phi) \equiv \t\herm{_\a_b} \t\varphi{^\a} \t\phi{^b}$.
As a consequence, the Klein–Gordon products
\begin{align}
	\pair{\varphi, \phi}
		= \int_\mfd \t j{^\mu}[\varphi, \phi] \t{\d\mfd}{_\mu}
\end{align}
are time-independent in the sense of evaluating to the same number for
all Cauchy surfaces $\mfd$.  Explicitly, in a stationary metric of the
form \eqref{eq:metric decomposition}, one has
\begin{align}
	\label{eq:KG product coordinates}
	\hspace{-2em}
	\pair{\varphi, \phi}
		= \frac{\i}{\hbar} \int \hspace{-0.35em} \t\Metric{^0^\mu} [
			\herm(\t\D{_\mu} \varphi, \phi)
			- \herm(\varphi, \t\D{_\mu} \phi)
		] \lapse \sqrt{\metric} \, \d^3 x.
\end{align}
As is explained below, the Klein–Gordon product plays a central role
in the quantum theory.

\section{Quantization}

In quantum field theory, “basic observables” form an involutive unital
algebra generated by expressions of the form $\Phi(\varphi)$,
$\Phi^\dagger(\phi)$, subject to the relations
\begin{subequations}
\begin{align}
	\label{eq:AQFT linearity}
	\Phi^\dagger(\phi_1 + \lambda \phi_2)
		&= \Phi^\dagger(\phi_1) + \lambda \Phi^\dagger(\phi_2),
	\\
	\label{eq:AQFT involution}
	\Phi(\varphi)^\dagger
		&= \Phi^\dagger(\bar\varphi),
	\\
	\label{eq:AQFT Heisenberg}
	[\Phi(\varphi), \Phi^\dagger(\phi)]_\varepsilon
		&= \pair{\varphi, \phi } e,
\end{align}
\end{subequations}
where $[A, B]_\varepsilon = A B + \varepsilon B A$ is the commutator
($\varepsilon = -1$) or anti-commutator ($\varepsilon = +1$) and $e$
denotes the unit element of the algebra
\cite[Sec.~2.1]{2015PhR...574....1H}.  Heuristically, $\Phi(\varphi)$,
$\Phi^\dagger(\phi)$ can be interpreted as expressions arising from
taking Klein–Gordon products of classical solutions $\varphi$ and
$\phi$ with quantum field operators $\hat\chi$ and
${\hat\chi}^\dagger$ according to the formulas
\begin{align}
	\Phi(\varphi)
		&= \pair{ \varphi, \hat\chi },
	&
	\Phi^\dagger(\phi)
		&= \pair{ \hat\chi^\dagger, \phi }.
\end{align}
This form immediately suggests \cref{eq:AQFT linearity,eq:AQFT
  involution}, while \cref{eq:AQFT Heisenberg} encodes Heisenberg’s
equal-time commutation relations.  However, due to the mathematical
difficulties in rigorously defining $\hat\chi$ and
${\hat\chi}^\dagger$, it is customary to work with $\Phi(\varphi)$,
$\Phi^\dagger(\phi)$ instead \cite{Wald_1994}. These expressions can
be understood either in terms of an abstract involutive algebra
(constructed from the free algebra over the space of $\varphi$ and
$\phi$ by taking appropriate quotients), or by an explicit Fock
construction in stationary space-times, see, e.g., Sec.~3.4.2 of
Ref.~\cite{Derezinski_Gerard_2022}.  Specifically, let $\solns$ and
$\bar\solns$ denote the spaces of all fields $\phi$ and $\varphi$
satisfying the field equations and having suitable decay properties,
respectively, and chose projectors $P_\pm$ onto subspaces such that
$\pair{ \barphi, \phi } \geq 0$ for all
$\phi \in \solns_+ \equiv P_+ \solns$ and
$\pair{ \barphi, \phi } \leq 0$ for all
$\phi \in \solns_- \equiv P_- \solns$ [in stationary space-times, such
projectors are typically associated with positive or negative
frequencies]. After factoring out solutions of vanishing norm and
taking closures to obtain a Hilbert space, one can construct
(depending on $\varepsilon$) symmetric or anti-symmetric Fock spaces
$\Phi_+$ and $\Phi_-$ with associated annihilation operators $a$ on
$\Phi_+$ (describing particles) and $b$ on $\Phi_-$ (describing
anti-particles).  Taking tensor products and identifying, for
notational simplicity, $a$ with $a \otimes \mathrm I$ and $b$ with
$\mathrm I \otimes b$, and similarly for $a^\dagger$ and $b^\dagger$,
the smeared field operators can be defined as
\begin{subequations}
\begin{align}
	\Phi(\varphi)
		&= a(P_- \varphi) - b(P_+ \varphi)^{\phantom\dagger},
	\\
	\Phi^\dagger(\phi)
		&= a^\dagger(P_+ \phi) - b^\dagger(P_- \phi).
\end{align}
\end{subequations}
Denoting the Fock vacuum by $\ket 0$ (using Dirac notation), the
transition amplitude for two (potentially unnormalized)
single-particle state vectors $\ket{\phi'} = a^\dagger(\phi') \ket 0$
and $\ket{\phi''} = a^\dagger(\phi'') \ket 0$, with both
$\phi', \phi'' \in \solns_+$, is given by
\begin{multline}
	\label{eq:transition amplitude}
	\hspace{-2em}
	\bbraket{\phi' | \phi''}
		=
		\frac{\braket{0|a(\barphi') a^\dagger(\phi'')|0}}{\sqrt{\braket{0|a(\barphi') a^\dagger(\phi')|0} \! \braket{0|a(\barphi'') a^\dagger(\phi'')|0}}}
		\\= 
		\frac{\pair{ \barphi', \phi'' }}{\sqrt{ \pair{ \barphi', \phi' } \pair{ \barphi'', \phi'' }}}.
\end{multline}
The general quantization scheme described here does not require
“complete sets” of explicit mode solutions. Using \cref{eq:transition
  amplitude}, the task of computing transition amplitudes is reduced
to the problem of computing Klein–Gordon products of specific
(experimentally relevant) solutions. Perturbative calculations of
these products are presented below.

\section{Quasi-Newtonian limit}
\label{supp:s:quasi-Newtonian limit}

The quasi-Newtonian approximation is obtained by taking the formal
limit $c \to \infty$, in which $\phi$ is written as
\begin{align}
	\phi
		&= \frac{\hbar}{\sqrt{2 m}} [\psi_\QN + O(1/c^2)]  \e^{-\ii m c^2 t / \hbar},
\end{align}
and the energy operator $\E$, lapse $\lapse$, and shift $\t\shift{_i}$
take the form
\begin{subequations}
\begin{align}
	\E &= m c^2 I + \H_\QN + O(1/c^2),
	\\
	\lapse &= 1 + \llapse / c^2 + O(1/c^4),
	&
	\\
	\t\shift{_i} &= \t v{_i} /c + O(1/c^3).
\end{align}
\end{subequations}
Here, $I$ denotes the unit matrix, and the overall factor
$\hbar / \sqrt{2m}$ ensures that $\psi_\QN$ has dimension
$1/\sqrt{\dimf L^3}$.  At leading order in $1/c^2$, this yields a
Schrödinger equation of the form
\begin{multline}
	\ii \hbar \t\p{_t} \psi_\QN
		= \frac{1}{2m} \t\metric{^i^j} (\t{\hat p}{_i} - m \t v{_i}) (\t{\hat p}{_j} - m \t v{_j}) \psi
		\\
		+ m \llapse \psi
		+ q V \psi
		+ \H_\QN \psi,
\end{multline}
where $\t{\hat p}{_i} = -\ii \hbar \t\del{_i} - q \t A{_i}$ is the
spatial momentum operator, with $\t\del{_i}$ denoting the Levi-Civita
derivative associated to the spatial metric $\t\metric{_i_j}$, and
$V = - \t A{_t}$ is the scalar electromagnetic potential.

The quasi-Newtonian limit considered here generalizes the more
commonly studied Newtonian limit by allowing for a non-flat spatial
metric $\t\metric{_i_j}$ and a non-vanishing velocity field
$\t v{_i}$. Setting $\t\metric{_i_j} = \t\delta{_i_j}$ and
$\t v{_i} = 0$ recovers the standard Schrödinger equation for a
particle of mass $m$, charge $q$, and internal Hamiltonian $\H_\QN$,
while setting $\t\metric{_i_j} = \t\delta{_i_j}$,
$\t v{_i} = \t\varepsilon{_i_j_k} \t\Omega{^j} \t x{^j}$ and
$\llapse = - \half \Omega^2 |x|^2$ recovers the Schrödinger equation
in a uniformly rotating frame \cite{2003quant.ph..5081A}.

Irrespective of the explicit form of $\t v{_i}$ and $\t\metric{_i_j}$,
the Klein–Gordon products reduce to $L^2$-products of the form
\begin{align}
	\hspace{-2em}
	\pair{ \barpsi', \psi'' }_\QN
	= 
	\lim_{c \to \infty}
	\pair{ \barphi', \phi'' }
		= \int_\mfd\! \herm(\barpsi', \psi'') \sqrt{\metric} \, \d^3 x,
\end{align}
where $\phi'$ and $\phi''$ are solutions to the \emph{exact} field
equation with the asymptotic behavior
$\phi \sim (\hbar / \sqrt{2m}) \psi \e^{-\i m c^2 / \hbar}$ as
$c \to \infty$.  A proof of the existence of such solutions is not
attempted here (this is a question of \PDE\ analysis).

Overall, the model described here can be regarded as a generalization
of the Schrödinger equation, with a finite number of \IDOFs, to curved
space-times.

\section{Semi-classical limit}
\label{supp:s:semi-classical limit}

The semi-classical approximation consists in a formal limit
$\hbar \to 0$, where the field $\phi$ is expanded as
\begin{align}
	\phi
		= \frac{\hbar}{\sqrt{2m}}[\chi_\SC + \hbar \chi_\SC' + O(\hbar^2)] \e^{\ii S / \hbar},
\end{align}
where $m$ is a constant mass parameter (such as a lower bound on the
eigenvalues of $\E$), and the energy matrix is written as
\begin{align}
	\E
		= \E_\SC
		+ \hbar \E_\SC'
		+ O(\hbar^2).
\end{align}
At leading and next-to-leading order in $\hbar$, this results in the
equations
\begin{align}
	\label{s:eq:SC 0}
	P \chi_\SC &=
		0,
	\\
	\label{s:eq:SC 1}
	\begin{split}
		P \chi'_\SC &=
			- 2 \ii \t\Metric{^\mu^\nu} \t p{_\mu} \t\Del{_\nu} \chi_\SC
			- \ii (\t\Metric{^\mu^\nu} \t\Del{_\mu} \t p{_\nu}) \chi_\SC
			\\&\quad
			+ c^{-2} \{\E_\SC, \E_\SC'\} \chi_\SC,
	\end{split}
\end{align}
where $\{\cdot, \cdot\}$ denotes the anti-commutator and $P$ is the
principal symbol
\begin{align}
	P
		&= - \t\Metric{^\mu^\nu} \t p{_\mu} \t p{_\nu} I - \E_\SC^2 / c^2,
\end{align}
in which $\t p{_\mu} = \t\Del{_\mu} S - q \t A{_\mu}$.  Assuming all
eigenvalues of $\E_\SC$ to be positive, \cref{s:eq:SC 0} implies that
$\chi_\SC$ is an eigenvector of $\E_\SC$.  Denoting by $E$ the
corresponding eigenvalue (which generically depends on space and time
whenever $\E_\SC$ does), one has
\begin{align}
	- \t\Metric{^\mu^\nu} \t p{_\mu} \t p{_\nu} = (M c)^2,
\end{align}
where $M = E/c^2$.  As a consequence, $S$ satisfies the
Hamilton–Jacobi equation
\begin{align}
	- \t\Metric{^\mu^\nu} (\t\Del{_\mu} S - q \t A{_\mu}) (\t\Del{_\nu} S - q \t A{_\nu})
		= (M c)^2,
\end{align}
which can be solved using characteristics, i.e., integral curves of
the four-velocity $\t u{^\mu} = \t\Metric{^\mu^\nu} \t p{_\nu} / M$.
Using the explicit form of $\t p{_\mu}$ and
$\t\Metric{_\nu_\rho} \t u{^\nu} \t u{^\rho} = -c^2$ (which implies
$\t\Metric{_\nu_\rho} \t u{^\nu} \t\Del{_\mu }\t u{^\rho} = 0$), the
four-force along these curves can be computed as follows
\begin{align}
	\begin{split}
		\t f{_\mu} &
			= \t u{^\nu} \t\Del{_\nu} \t p{_\mu}
			\\&
			= \t u{^\nu} \t\Del{_\mu} \t p{_\nu} - (\t\Del{_\mu} \t p{_\nu} - \t\Del{_\nu} \t p{_\mu}) \t u{^\nu}
			\\&
			= \t u{^\nu} \t\Del{_\mu} (M \t\Metric{_\nu_\rho} \t u{^\rho} ) + q (\t\Del{_\mu} \t A{_\nu} - \t\Del{_\nu} \t A{_\mu}) \t u{^\nu} 
			\\&
			= \t\Metric{_\nu_\rho} \t u{^\nu} \t u{^\rho} \t\Del{_\mu} M  + q \t F{_\mu_\nu} \t u{^\nu} 
			\\&
			= - c^2 \t\Del{_\mu} M + q \t F{_\mu_\nu} \t u{^\nu} 
			\\&
			= - \t\Del{_\mu} E + q \t F{_\mu_\nu} \t u{^\nu},
	\end{split}
\end{align}
where $\t F{_\mu_\nu} = \t{(\dd A)}{_\mu_\nu}$ is the electromagnetic
field strength (Faraday tensor).  The four-acceleration
$\t a{^\mu} = \t u{^\nu} \t\nabla{_\nu} \t u{^\mu}$ then takes the
form
\begin{align}
	\t a{^\mu}
		= \tfrac{1}{M} [\t\Metric{^\mu^\nu} + \t u{^\mu} \t u{^\nu}/c^2] \t f{_\nu}.
\end{align}
Along such curves, $S$ evolves according to the equation
\begin{align}
	\label{s:eq:dS/dtau}
	\t u{^\mu} \t\Del{_\mu} S
		= - M c^2 + q \t A{_\mu} \t u{^\mu}.
\end{align}
Hence, the phase $S/\hbar$ can be computed by integrating
\cref{s:eq:dS/dtau} along rays satisfying the classical equations of
motion.

So far, \cref{s:eq:SC 0} was primarily used to determine the phase
function $S/\hbar$, with the only requirement on $\chi_\SC$ demanding
it to be an eigenvector of $\E_\SC$ with eigenvalue $E$ (which is
assumed to vary smoothly with space and time).  In general, the
eigenspace of $\E_\SC$ corresponding to the eigenvalue $E$ may be
multidimensional, hence further analysis is needed to determine the
evolution of the remaining degrees of freedom in $\chi_\SC$.  The
relevant equation can be obtained from \cref{s:eq:SC 1}. The key
observation here is that the principal symbol $P$ is not invertible,
and hence the right-hand side of \cref{s:eq:SC 1} must lie in the same
eigenspace as $\chi_\SC$ in order for this equation to be solvable in
terms of $\chi'_\SC$.  Denoting by $\Riesz$ the orthogonal projector
onto the eigenspace of $\E_\SC$ associated to the eigenvalue $E$, this
condition can be written as
\begin{align}
	\hspace{-1em}
	\ii \t u{^\mu} \Riesz \t\Del{_\mu} \chi_\SC
	= -\tfrac{\i}{2 M} (\t\Metric{^\mu^\nu} \t\Del{_\mu} \t p{_\nu}) \chi_\SC
	+ \Riesz \E'_\SC \chi_\SC.
\end{align}
Here, $\chi_\SC$ need not be normalized.  Indeed, writing
$\chi_\SC = \A \psi_\SC$, where $\psi_\SC$ is normalized to
$\herm(\bar \psi_\SC, \psi_\SC) = 1$, one obtains the equivalent
system
\begin{subequations}
\begin{align}
	\t u{^\mu} \t\Del{_\mu} \A
		&= - \tfrac{1}{2 M} (\t\Metric{^\mu^\nu} \t\Del{_\mu} \t p{_\nu}) \A,
	\\
	\ii \t u{^\mu} \Riesz \t\Del{_\mu} \psi_\SC
		&=  \Riesz \E'_\SC \psi_\SC.
\end{align}
\end{subequations}
The transport equation for $\psi_\SC$ can be written in the equivalent
form
\begin{align}
	\ii \t u{^\mu} \t\bD{_\mu} \psi_\SC
		=  \E'_\SC \psi_\SC +  [\Riesz, \E'_\SC] \psi_\SC
		,
\end{align}
where $\t\bD{_\mu}$ is the Berry connection acting as
\begin{align}
	\t\bD{_\mu} \psi_\SC
		= \Riesz \t\Del{_\mu} \psi_\SC
		= \t\Del{_\mu} \psi_\SC - (\t\Del{_\mu} \Riesz) \psi_\SC.
\end{align}
Explicitly, decomposing $\psi_\SC$ as
$\t*\psi{_\SC} = \t*\psi{_\SC^\alpha} \t\sigma{_\alpha}$, where, at
each point, the local frame $\t\sigma{_\alpha}$ spans the eigenspace
$\E_\SC$ with eigenvalue $E$, one has
\begin{align}
	\t\bD{_\mu} \t*\psi{_\SC}
		= [
			\t\p{_\mu} \t*\psi{_\SC^\alpha}
			- \ii \t\varpi{^\alpha_\beta_\mu} \t*\psi{_\SC^\beta}
		] \t\sigma{_\alpha},
\end{align}
in which $\t\varpi{^\alpha_\beta_\mu}$ are Berry connection
coefficients of the first kind.  These are related to the connection
coefficients of the second kind,
$\t\varpi{_\balpha_\beta_\mu} = \ii \herm(\t{\bar\sigma}{_\balpha},
\t\Del{_\mu} \t\sigma{_\beta})$, by
$\t\varpi{_\balpha_\beta_\mu} = \herm(\t{\bar\sigma}{_\balpha},
\t\sigma{_\gamma}) \t\varpi{^\gamma_\beta_\mu}$.  In particular,
$\t\varpi{^\alpha_\beta_\mu}$ and $\t\varpi{_\balpha_\beta_\mu}$
coincide if the frame $\t\sigma{_\alpha}$ is orthonormal.  Using this
notation, the transport law can be written in the form of a
Schrödinger equation
\begin{align}
	\ii \hbar \t u{^\mu} \t\p{_\mu} \t*\psi{_\SC^\alpha}
		= \t{{(\t*\H{_\SC})}}{^\alpha_\beta} \t\psi{_\SC^\beta}
		- \hbar \t\varpi{^\alpha_\beta_\mu} \t*\psi{_\SC^\beta}  \t u{^\mu},
\end{align}
where $\H_\SC = \hbar \E'_\SC$ is the internal Hamiltonian.

The Klein–Gordon product of such (approximate) solutions is given by
the asymptotic behavior the inner product $\pair{\barphi', \phi''}$,
as $\hbar \to 0$, where $\phi'$ and $\phi''$ are solutions to the
exact field equations with the asymptotic behavior
$\phi \sim (\hbar / \sqrt{2m}) \A \psi \e^{\i S/\hbar}$ (similarly to
the quasi-Newtonian limit, a rigorous existence proof is beyond the
scope of this letter).  Using the explicit form of the product in
\cref{eq:KG product coordinates} and the principle of stationary
phases, see, e.g., Sec.~7.7 of Ref.~\cite{Hörmander1}, shows that the
product vanishes if the relative phase $\varDelta S = S' - S''$ has no
stationary points within the support of
$\bar\A' \A'' \herm(\barpsi', \psi'')$.  Alternatively, if
$\varDelta S$ has stationary points, Theorem 7.7.6 of
Ref.~\cite{Hörmander1} provides an asymptotic expansion for the case
when the Hessian of $\varDelta S$ is non-degenerate (involving a sum
over all critical points). The simplest case, however, occurs when
$\varDelta S$ is constant, for one then has
\begin{multline}
	\pair{\A', \psi', S' | \A'', \psi'', S''}_\SC
	\\=
	\e^{- \ii \varDelta S / \hbar} \! \int_\Sigma \hspace{-0.5em} \bar\A' \A'' \herm(\barpsi', \psi'') \varrho \sqrt{\metric} \d^3 x,
\end{multline}
with $\varrho = \lapse \t\Metric{^0^\mu} \t p{_\mu} / m$.  As stated,
this equation requires adapted coordinates so that $\Sigma$ is a
level-set of the temporal coordinate $t$ (hence
$\t n{_\mu} = \t\p{_\mu} t$ is co-normal to $\Sigma$), and $\t\p{_t}$
is a Killing vector field.  According to \cref{eq:metric
  decomposition}, the corresponding four-velocity of stationary
observers is $\mathring u = \lapse^{-1} \t\p{_t}$.  This implies
$\t{\mathring u}{^\mu} \t n{_\mu} = 1/\lapse$ and
$\t\Metric{^0^\mu} \t p{_\mu} = M \t u{^\mu} \t n{_\mu}$, where
$\t u{^\mu}$ is the four-velocity corresponding to the four-momentum
$\t p{_\mu}$. As a consequence, the weight $\varrho$ admits the
covariant expression
\begin{align}
	\varrho = \frac{M}{m} \frac{\t u{^\mu} \t n{_\mu}}{\t{\mathring u}{^\nu} \t n{_\nu}}.
\end{align}
%


\bibliography{bibliography}